\documentclass[twocolumn,showpacs]{revtex4}

\begin{document}

%%%%%%%%%%%%%%%%%%%%%%%%%%%%%%%%%%%%%%%%%%%%%%%%%%%%%%%%%%%%%%%%
%%% My Defs:
% \newcommand{\ftimes}{f\times}   %  <-- use Atten. factor f
  \newcommand{\ftimes}{}          %  <-- get rid of Atten. factor f

\preprint{AEI-2002-048,UTBRG-2002-06}

\title{Binary black hole initial data for numerical general relativity based on
post-Newtonian data}

\author{Wolfgang Tichy$^{1}$, Bernd Br\"ugmann$^{1}$, 
Manuela Campanelli$^{1,2}$, Peter Diener$^{1}$}
\affiliation{$1$ 
Albert-Einstein-Institut, Max-Planck-Institut f{\"u}r
Gravitationsphysik, Am M\"uhlenberg 1, D-14476 Golm, Germany \\
$2$ Department of Physics and Astronomy, 
The University of Texas at Brownsville, Brownsville, Texas 78520}
\date{July 1, 2002}

\input epsf

\begin{abstract}
With the goal of taking a step toward the construction of
astrophysically realistic initial data for numerical simulations of
black holes, we for the first time derive a family of fully general
relativistic initial data based on post-2-Newtonian expansions of the
3-metric and extrinsic curvature without spin. It is expected that
such initial data provide a direct connection with the early inspiral
phase of the binary system.  We discuss a straightforward numerical
implementation, which is based on a generalized puncture
method. Furthermore, we suggest a method to address some of the
inherent ambiguity in mapping post-Newtonian data onto a solution of
the general relativistic constraints.
\end{abstract}

\pacs{04.25.Dm, 04.25.Nx, 04.30.Db, 04.70.Bw}

\maketitle

\section{Introduction}

One of the most exciting scientific objectives of gravitational wave
astronomy involves the search for and detailed study of signals from
sources that contain binary black holes. Mergers of two black holes
both with masses of $\sim 10-100M_{\odot}$ will be observable by the
ground based gravitational wave detectors, such as GEO600, LIGO and
others~\cite{Schutz99}. These systems are highly relativistic once
they enter the sensitive frequency band ($\sim 50 - 200 \, {\rm Hz}$)
of the detector. For LISA, gravitational waves from super-massive
binary black hole mergers (e.g.\ black holes with mass greater than 
$10^6M_\odot$) are very strong, with high signal-to-noise ratios up to
$10^4$~\cite{Hughes:2001ch}, making these events observable from
almost anywhere in the universe. Astrophysically realistic models of
binary black hole coalescence are therefore required to study these
phenomena in detail~\cite{Flanagan97a}.

To solve the full Einstein equations in the dynamic, non-linear phase
at the end of the binary black hole inspiral we turn to numerical
relativity.  Numerical relativity has advanced to the point where a
time interval of up to $40M$ (where $M$ is the total mass) of the
merger phase of two black holes can be computed if the black holes
start out close to each other
\cite{Bruegmann97,Brandt00,Alcubierre00b}. Recent simulations of
head-on collisions of black holes last significantly longer and give
reason for optimism for the orbiting case~\cite{Alcubierre02a}. An
approach to produce at least moderately accurate models for the wave
forms generated in binary black hole mergers was recently developed in
the so-called Lazarus
project~\cite{Baker00a,Baker00b,Baker:2001sf,Baker:2001nu,Baker:2002qf},
a technique that bridges `close' and `far' limit approximations with
full numerical relativity. This approach
has lead to the first approximate theoretical estimates for the
gravitational radiation wave forms and energy to be expected from the
plunge of orbiting non-spinning binary black holes to
coalescence~\cite{Baker:2001nu,Baker:2002qf}.

Due to theoretical and numerical limitations, all current numerical
simulations must begin by specifying initial data when the black holes
are already very close (separation $\lesssim 7M$). There is
a push to place the starting point of these simulations at earlier
times, say at a few orbits before a fiducial innermost stable circular
orbit (ISCO) which approximately marks the transition from the
inspiral phase to the plunge and merger.  But whatever the starting
point, the simulation will only be astrophysically meaningful if
it starts with astrophysically realistic initial data.

The question we want to address in this paper is therefore how to obtain
astrophysically realistic initial data for numerical simulations of
binary black hole systems. 
In general relativity the initial data must fulfill constraint
equations, so only part of the data are freely specifiable, and the
rest is determined by solving the constraint equations (for a review
see e.g.~\cite{Cook:2000}). A lot of the work in constructing initial
data has focused on approaches that pick the freely specifiable part
of the data with the aim of simplifying the constraint equations,
rather than using astrophysically realistic initial data. A standard
assumption is that the 3-metric is conformally flat and the extrinsic
curvature is derived from a purely longitudinal ansatz (see e.g.\
\cite{Cook:2000,Bowen80,Cook94,Brandt97b}).  Currently, there are a
number of new approaches
~\cite{Marronetti00,Grandclement:2001ed,Cook:2001wi,Dain00,Dain:2002ee} to
specify `improved', including non-conformally flat, initial data for
binary black holes.

However, none of these approaches to construct initial data makes
explicit use of information from an approximation procedure such as
the post-Newtonian (PN) method, which is believed to accurately
represent astrophysical systems in the limit of slow-moving/far-apart
black holes.  An approximate binary black hole metric based on
post-1-Newtonian (1PN) information in a corotating gauge has been
derived by Alvi~\cite{Alvi99}. However, at present this metric cannot
be used in numerical simulations due to the presence of
discontinuities in the matching regions~\cite{Jansen_priv}. An
interesting approach based on quasi-equilibrium sequences of initial
data has been studied numerically, e.g.~\cite{Duez00}, although some
aspects of the method appear to be based on Newtonian or 1PN
assumptions.

In this paper we describe a method to generate new fully general
relativistic initial data for two inspiraling black holes from PN
expressions. The motivation for this method is that even though PN
theory may not be able to evolve two black holes when they get close,
it can still provide initial data for fully nonlinear numerical
simulations when we start at a separation where PN theory is valid.
In particular, we obtain an explicit far limit interface for the
Lazarus approach.  Our method allows us to incorporate
information from the PN treatment and should eventually provide a
direct connection to the inspiral radiation.

Like in other approaches, we start from expressions for the 3-metric
and extrinsic curvature in a convenient gauge. We use expressions for
the 3-metric and its conjugate momentum up to PN order $(v/c)^5$,
computed in the canonical formalism of ADM by Jaranowski and
Sch\"afer~\cite{Jaranowski98a}. This order corresponds to 2.5PN in the
3-metric and 2PN in the conjugate momentum, since the latter contains
a time derivative. Therefore, the PN data are accurate to 2PN.

The 3-metric and its conjugate momentum are derived
together with a two-body Hamiltonian using coordinate
conditions~\cite{Ohta74,Schaefer85,Damour:2001bu}, which correspond to
the ADM transverse-traceless (ADMTT) gauge. Note that there are
several other formulations and gauges for PN theory, see
e.g.~\cite{Blanchet02} for a review.
The ADMTT gauge has several advantages:
(i) we can easily find
expressions for 3-metric and extrinsic curvature,
(ii) unlike in the harmonic gauge no logarithmic divergences appear,
(iii) for a single black hole the data simply reduce to 
Schwarzschild in standard isotropic coordinates,  
(iv) up to $(v/c)^3$ the data look like in the puncture approach
\cite{Brandt97b}, which simplifies calculations, and
(v) the trace of the extrinsic curvature vanishes up to order $(v/c)^6$,
so that we can set it to zero (if we go only up to order $ (v/c)^5$),
which can be used to decouple the Hamiltonian 
constraint equation from the momentum constraint equations. 
In the ADMTT gauge the 3-metric is conformally flat up to order $ (v/c)^3$,
at order $ (v/c)^4$ deviations from conformal flatness enter. 
The extrinsic curvature up to order $ (v/c)^3$ is simply of 
Bowen-York form \cite{Bowen80}, with correction terms of order $ (v/c)^5$.

We will use the York-Lichnerowicz conformal decomposition
\cite{York73} and use the PN data as the freely specifiable data. We
numerically solve for a new conformal factor $\Psi$ and the usual
correction to the extrinsic curvature, given by a vector
potential $W^i$. The new extrinsic curvature and the 3-metric multiplied by
$\Psi^4$ are then guaranteed to fulfill the constraints.  The real
problem in this approach is to find a numerical scheme which can deal
with the divergences in the PN data at the center of each black hole.
The most serious divergence occurs in the PN conformal factor
$\psi_{PN}$ of the conformally flat part of the 3-metric.  We
therefore rescale the PN data by appropriate powers of $\psi_{PN}$ to
generate a well behaved 3-metric. If we then use the conformally
rescaled data as the freely specifiable data and make the ansatz that
the new conformal factor $\Psi$ is the PN conformal factor $\psi_{PN}$
plus a finite correction $u$, we arrive at elliptical equations which
can be solved numerically. The splitting of the new conformal factor
into $\Psi = \psi_{PN} + u $ is very similar to the puncture approach
\cite{Brandt97b}, except that in our case the momentum constraint has
to be solved numerically as well.

Let us point out several issues that arise in the construction of
solutions to the constraints of the full theory based on PN data.
First of all, the accuracy of the PN approximation increases with
the separation of the binary, and the same is therefore true for the
numerical data. Second, PN theory typically deals with point
particles rather than black holes. One has to somehow introduce black
holes into the theory, which leads to a certain arbitrariness of the
data near the black holes. We make the specific choice contained
in~\cite{Jaranowski98a}. Note that since we are solving elliptic equations,
the data near the black holes affect the solution everywhere.
%Third, since the data are valid only for
%$(v/c)^2 \sim M/r \ll 1$, it is clear they are not valid close to the
%particles. 
Third, some of the PN expressions that we use are near zone
expansions which are invalid far from the particles. This means we
have data only in a limited region of space.

Finally, the reader should be aware of the following basic feature of
the York procedure to compute initial data.
Given valid free data, which in our case is derived from the PN data,
the procedure projects the data onto the solution space of the
constraints. This projection maps the PN data somewhere, but is the end
point better than the starting point? We have to make sure that 
we do not loose the advantage of starting with PN data over, say,
simply using PN orbital parameters in the conformally flat data approach.
After describing and resolving several technical issues in the
construction of our data set, we will therefore (i) quantify
the `kick' from PN to fully relativistic data, and (ii) suggest
a concrete method for improving the results of our straightforward 
first implementation.

\subsection{Notation}

We use units where $G=c=1$. Lowercase Latin indices denote the spatial
components of tensors. 
The coordinate locations of the two
particles are denoted by $(x_1, y_1, z_1)$ and $(x_2, y_2, z_2)$. We
define
\begin{equation}
r_{A} := \sqrt{(x-x_{A})^2 + (y-y_{A})^2 + (z-z_{A})^2} .
\end{equation}
and 
\begin{equation}
n_{A}^i := (x-x_{A}, y-y_{A}, z-z_{A})/r_{A} ,
\end{equation}
where the subscript $A$ labels the particles. Furthermore we introduce
\begin{equation}
r_{12} := \sqrt{(x_1-x_2)^2 + (y_1-y_2)^2 + (z_1-z_2)^2}
\end{equation}
to denote the separation between the particles.
All terms carrying a superscript $TT$ are
transverse traceless with respect to the flat 3-metric $\delta_{ij}$.

\section{The PN expressions for 3-metric and extrinsic curvature}
\label{sec-expr_metric_curv}

Our starting point is the expressions for the PN 3-metric
$g^{PN}_{ij}$ and the PN 3-momentum $\pi_{PN}^{ij}$ computed in the
ADMTT gauge \cite{Jaranowski98a}. The ADMTT gauge is specified by
demanding that the 3-metric has the form
\begin{equation}
\label{ADM_1}
g^{PN}_{ij} = \psi_{PN}^4 \delta_{ij} + h_{ij }^{TT} .
\end{equation}
and that the conjugate momentum fulfills 
\begin{equation}
\label{ADM_2}
\pi_{PN}^{ij} \delta_{ij} = 0 .
\end{equation}
We explicitly include the formal PN expansion parameter $\epsilon \sim
v/c$ in all PN expressions, a subscript in round brackets will denote
the order of each term. When a PN term is evaluated numerically,
$\epsilon$ is set to one.

We start with the PN expression for the 3-metric \cite{Jaranowski98a} 
\begin{equation}
\label{gij_down}
g^{PN}_{ij} = \psi_{PN}^4 \delta_{ij} + \epsilon^4 h_{ij (4)}^{TT} 
                + \epsilon^5 h_{ij (5)}^{TT} + O(\epsilon^6) ,
\end{equation}
where the conformal factor of PN theory is given by
\begin{equation}
\label{psi_PN_of_phi}
\psi_{PN}=1     
      +\frac{1}{8} \left(\epsilon^2 \phi_{(2)} + \epsilon^4 \phi_{(4)}\right)
        +O(\epsilon^6) .
\end{equation}
Using the expressions for $\phi_{(2)}$ and $\phi_{(4)}$ 
given in \cite{Jaranowski98a} we see that the conformal factor $\psi_{PN}$ can
be written in the simple form
\begin{equation}
\label{psiPN}
\psi_{PN}=1 +  \sum_{A=1}^2 \frac{E_A}{2r_A} +O(\epsilon^6) , 
\end{equation}
where the constants $E_1$ and $E_2$
depend only on the masses $m_1$, $m_2$, the momenta $p_1$, $p_2$ and
the separation $r_{12}$
of PN theory. They are given by
\begin{equation}
\label{EnergyInPsi}
E_{A} = \epsilon^2 m_{A} 
     +\epsilon^4 \left(\frac{p_{A}^2}{2m_{A}} 
                        -\frac{m_1 m_2}{2r_{12}}  \right)
\end{equation}
and can be regarded as the energy of each particle.

Note that the PN 3-metric is singular at the location of each
particle, since $\phi_{(2)} $, $\phi_{(4)}$ and $h_{ij (4)}^{TT}$ all
go like $\sim 1/r_A $ as particle $A$ is approached, and
$h_{ij(5)}^{TT}$ is regular. This means that the strongest
singularity is in $\psi_{PN}^4 \sim 1/r_A^4$ and that the
$\psi_{PN}^4$ term dominates near each particle. Hence near each
particle the 3-metric can be approximated by
\begin{equation}
g^{PN}_{ij} \approx \left( 1+ \frac{E_A}{2r_A}\right)^4 
                        \delta_{ij} +O(1/r_{A}^3)  ,
\label{gPNunexpanded}
\end{equation}
which is just the Schwarzschild 3-metric in isotropic coordinates.
For $r_A\rightarrow 0$ we approach the coordinate singularity that
represents the inner asymptotically flat end of Schwarzschild in
isotropic coordinates, which is also called the puncture
representation of Schwarzschild.  This shows that if we write the
3-metric as in Eq.~(\ref{gij_down}), we actually do have a black hole
centered on each particle. This is non-trivial since PN theory in
principle only describes particles.

On the other hand, if we expand the conformal factor in
Eq.~(\ref{gij_down}), the puncture singularity of Schwarzschild is no
longer present.
If we insert Eq.~(\ref{psi_PN_of_phi}) into Eq.~(\ref{gij_down}) and
expand in $\epsilon$ we obtain
\begin{eqnarray}
g^{PN }_{ij} &=& \left[1
+ \epsilon^2 \frac{1}{2}\phi_{(2)} 
+ \epsilon^4 \left(\frac{1}{2}\phi_{(4)} +\frac{3}{32}\phi_{(2)}^2 \right)
\right] \delta_{ij} \nonumber \\
&&+ \epsilon^4 h_{ij (4)}^{TT}
  + \epsilon^5 h_{ij (5)}^{TT} + O(\epsilon^6) ,
\end{eqnarray}
which goes like 
\begin{equation}
g^{PN}_{ij} \approx \left( \frac{\mbox{const}}{r_A^2}\right) 
                        \delta_{ij} +O(1/r_{A}) ,
\label{gPNexpanded}
\end{equation}
near each particle. One necessary condition for a black hole is the
presence of a marginally trapped surface, and while the Schwarzschild
metric in isotropic coordinates has a minimal surface at radius $M/2$,
the term in $1/r_A^2$ in (\ref{gPNexpanded}) leads to a minimum in
area at radius zero (ignoring the extrinsic curvature terms). 
Therefore the particle is not necessarily surrounded by a horizon.  

From now on we will use the 3-metric of~\cite{Jaranowski98a} as
written in Eq.~(\ref{gij_down}), without expanding $\psi_{PN}^4$ in
$\epsilon$, in order to make sure that we have black holes in our
data. The puncture coordinate singularity has replaced the point
particle singularity. This choice is somewhat ad hoc, but since PN
theory is not valid near the particles anyway, we have to make some
choice, and putting in black holes as punctures seems natural.

The determinant of $g^{PN}_{ij}$ is 
\begin{equation}
\label{detg}
g^{PN}=\psi_{PN}^{12} + O(\epsilon^6) ,
\end{equation}
since $\delta^{ij} h_{ij}^{TT}=0$. 

The PN expansion for the conjugate momentum is \cite{Jaranowski98a}
\begin{equation}
\label{pi_ij_PN}
\pi_{PN}^{ij} =  \epsilon^3 \tilde{\pi}^{ij}_{(3)}
                + \epsilon^5  \tilde{\pi}^{ij}_{(5)} 
                + \epsilon^5 \pi^{ij TT}_{(5)}
                 + O(\epsilon^6) , 
\end{equation}
where
\begin{equation}
\label{pi_ij_tilde5}
\tilde{\pi}^{ij}_{(5)} = -\frac{1}{2}\phi_{(2)}\tilde{\pi}^{ij}_{(3)}
                        +\frac{1}{2}(\phi_{(2)}\tilde{\pi}^{ij}_{(3)})^{TT}
\end{equation}
and
\begin{equation}
\label{pi_ij_TT5}
\pi^{ij TT}_{(5)} = \frac{1}{2}\dot{h}_{ij (4)}^{TT}
                        +\frac{1}{2}(\phi_{(2)}\tilde{\pi}^{ij}_{(3)})^{TT} .
\end{equation}  
As in the case of the 3-metric it turns out that $\pi_{PN}^{ij}$
in Eq.~(\ref{pi_ij_PN}) is singular, since 
$\tilde{\pi}^{ij}_{(3)}$, $\tilde{\pi}^{ij}_{(5)}$ and $\pi^{ij TT}_{(5)}$
all diverge at the location of each particle. But all these singularities
in $\pi_{PN}^{ij}$ up to $O(\epsilon^5)$ can be removed by rewriting 
Eq.~(\ref{pi_ij_PN}) as \cite{Schaefer_priv}
\begin{eqnarray}
\label{pi_ij_PN_resum}   
\pi_{PN}^{ij} &=& \psi_{PN}^{-4} \left[ \epsilon^3 \tilde{\pi}^{ij}_{(3)}
                + \epsilon^5 \frac{1}{2}\dot{h}_{ij (4)}^{TT}
                + \epsilon^5 (\phi_{(2)}\tilde{\pi}^{ij}_{(3)})^{TT}
                \right]
\nonumber \\ 
&& + O(\epsilon^6) .
\end{eqnarray}
which can be verified to agree with Eq.~(\ref{pi_ij_PN}) by
re-expanding $\psi_{PN}$ as in Eq.~(\ref{psi_PN_of_phi}) and keeping
only terms up to $O(\epsilon^5)$. Hence 
all singularities can be absorbed by the
conformal factor, which is the basis for the puncture method in
general \cite{Brandt97b,Alcubierre02a}.
%At first sight the disappearance of singularities in
%Eq.~(\ref{pi_ij_PN_resum}) may seem miraculous, but what really happened is
%that we added some terms of higher order in $\epsilon$ 
%to cancel the singularities, which drop out again if we re-expand
%Eq.~(\ref{pi_ij_PN_resum}).

Note that explicit expressions for $\phi_{(2)}$, $\phi_{(4)}$ 
and $\tilde{\pi}^{ij}_{(3)}$ can be found
in e.g.~\cite{Jaranowski98a} or \cite{Ohta74}. 
In addition Ohta et al~\cite{Ohta74}
also give an expression for the lapse up to $O(\epsilon^4)$
and for the shift up to $O(\epsilon^5)$.
The explicit expressions for $h_{ij (4)}^{TT}$, $\dot{h}_{ij (4)}^{TT}$,
and $\pi^{ij TT}_{(5)}$, however,
we obtained from Jaranowski and Sch\"afer in a Mathematica file.

It should also be noted that the analytic 
expressions \cite{Jaranowski98a} used for the
PN terms $\phi_{(2)}$, $\phi_{(4)}$ and $\tilde{\pi}^{ij}_{(3)}$ 
are valid everywhere,
while the expressions used for $h_{ij (4)}^{TT}$, $\dot{h}_{ij (4)}^{TT}$ 
and $h_{ij (5)}^{TT}$ are near zone expansions. 

The near zone expansion is valid only for $r \ll \lambda \sim \pi
\sqrt{r^3/(m_1+m_2)}$, where $r$ is the distance from the particle
sources and $\lambda$ is the wavelength.  In principle $h_{ij}^{TT}$
should be computed from a wave equation, but in the near zone this
equation can be simplified by replacing the d'Alembertian by a
Laplacian. This is exactly what Jaranowski and Sch\"afer
\cite{Jaranowski98a} do to arrive at the expression for $h_{ij}^{TT}$
we use.  In particular, the near zone expansion for $h_{ij (5)}^{TT}$
is a spatially constant tensor field that just varies in time. So for
the purpose of finding initial data it suffices to choose the initial
time such that $h_{ij (5)}^{TT}$ vanishes. Thus in all our numerical
computations we will set $h_{ij (5)}^{TT} = 0$.

Using the gauge condition (\ref{ADM_2}) we obtain
\begin{equation}
\label{Tracepi}
\pi_{PN} = g^{PN}_{ij}\pi_{PN}^{ij} = O(\epsilon^7) .
\end{equation}
The next task is to compute the extrinsic curvature 
\begin{equation}
K_{PN}^{ij} = -\frac{1}{\sqrt{g}}\left( \pi_{PN}^{ij} 
                                        - \frac{1}{2}\pi_{PN} g^{ij} \right)
\end{equation}
from the conjugate momentum $\pi_{PN}^{ij}$.
With the help of Eqs.~(\ref{detg}) and (\ref{Tracepi}), and using
the expressions for $\pi_{PN}^{ij}$ in Eq.~(\ref{pi_ij_PN_resum})
we find that the extrinsic curvature can be written as
\begin{eqnarray}
\label{Kij_up}
K_{PN}^{ij} &=& -\psi_{PN}^{-10} \left[ \epsilon^3 \tilde{\pi}^{ij}_{(3)} 
                + \epsilon^5 \frac{1}{2}\dot{h}_{ij (4)}^{TT}
                + \epsilon^5 (\phi_{(2)}\tilde{\pi}^{ij}_{(3)})^{TT} 
                \right] 
\nonumber \\
&& + O(\epsilon^6) ,
\end{eqnarray}
such that the conformal factor $\psi_{PN}$ is factored out.
The leading term in Eq.~(\ref{Kij_up}) is of Bowen-York form, i.e.
\begin{eqnarray}
-\tilde{\pi}^{ij}_{(3)} 
&=& \sum_{A=1}^2 \frac{3}{2 r_{A}^2}\left[ 
  p_A^i n_A^j +p_A^j n_A^i \right.
\nonumber \\
&& \quad
\left. -p_A^m n_A^n \delta_{mn}(\delta^{ij} -n_A^i n_A^j) 
  \right] .
\end{eqnarray}
Using that $\partial_i \tilde{\pi}^{ij}_{(3)}=0$ outside the
singularities and the fact that the last two terms 
inside the square bracket of Eq.~(\ref{Kij_up})
are transverse (with respect to $\delta_{ij}$), we find
\begin{equation}
\label{DivKij}
\partial_i (\psi_{PN}^{10} K_{PN}^{ij}) = O(\epsilon^6) 
\end{equation}
outside the singularities.
Moreover from Eq.~(\ref{Tracepi}) we have
\begin{equation}
\label{TraceK}
K_{PN}=g^{PN}_{ij} K_{PN}^{ij} = O(\epsilon^7)
\end{equation}
so that $K_{PN}^{ij}$ can be considered traceless up to $O(\epsilon^6)$.

\section{Circular orbits in PN theory}

The PN expressions given in section \ref{sec-expr_metric_curv} 
are valid for general orbits.
Any particular orbit is specified by giving the positions and momenta of the
two particles. In this paper we want to consider quasi-circular orbits,
since they are believed to be astrophysically most relevant.
For a given separation $r_{12}$ we therefore
choose the momenta $p_{A}^i$ such 
that we get a circular orbit of post-2-Newtonian (2PN) theory.
If we choose the center of mass to be at rest the two momenta
must be opposite in sign and equal in magnitude.
Also, for reasons of symmetry $p_{1}^i$ and $p_{2}^i$ for circular 
orbits must be perpendicular to the line connecting the two particles.
Next from the expressions for angular momentum and energy for circular 
orbits given by Sch\"afer and Wex \cite{Schaefer93}, 
we find that the momentum magnitude $p_{PN}^{circ}$
for circular orbits is given by
\begin{eqnarray}
\label{p_circ}
(p_{PN}^{circ})^2 &=& \mu^2 \frac{M}{r_{12}} 
     + \epsilon^2 4\mu^2 \frac{M^2}{r_{12}^2}  
     + \epsilon^4 (74 -43\frac{\mu}{M})\mu^2 \frac{M^3}{8r_{12}^3} 
\nonumber \\
&&     +O(\epsilon^{5}) , 
\end{eqnarray}
where $M=m_1+m_2$ and $\mu=m_1 m_2/M$.
If this formula for the momentum together with the separation
is inserted into the expressions for
3-metric and extrinsic curvature in section \ref{sec-expr_metric_curv},
we obtain PN initial data for circular orbits.
There are, however, at least two ways how this can be done.
One way is to always insert the momentum (\ref{p_circ})
to the highest order known, even in terms which are themselves 
say of $O(\epsilon^4)$. One might hope to thereby improve
the PN trajectory information in the initial data.
Another way is to consistently only keep terms up to a specified order,
say up to $O(\epsilon^5)$. As an example let us look at the PN conformal
factor given by Eqs.~(\ref{psiPN}) and (\ref{EnergyInPsi}). As one can see 
from Eq.~(\ref{EnergyInPsi}), the momentum terms are already $O(\epsilon^4)$,
so that if we insert Eq.~(\ref{p_circ}), we generate terms of 
$O(\epsilon^6)$ and $O(\epsilon^8)$, which should be dropped
if we consistently want to keep terms only up to $O(\epsilon^5)$.
We will see later that the ADM mass of the system is indeed sensitive to
whether or not we drop such terms in the conformal factor.

In order to compare with numerically computed 
ADM masses, we will also need an expression for PN total 
energy of the system. For circular orbits it is given by
\begin{eqnarray}
\label{PN_Etot}
E_{PN}^{circ}&=& M-\frac{\mu M}{2r_{12}}
\Bigg(1 + \epsilon^2 \Big[\frac{\mu}{M}-7\Big]\frac{M}{4r_{12}} \nonumber \\
      && \!\!\!\!\!\!\!\! 
+ \epsilon^4 \Big[-9+20\frac{\mu}{M}+\frac{\mu^2}{M^2}\Big]
                     \frac{M^2}{8r_{12}} \Bigg) 
                 + O(\epsilon^{6}) .
\end{eqnarray}

\section{Solving the constraints}

\subsection{The York Procedure}

The PN expressions for the 3-metric and the extrinsic curvature as
given in Eqs.~(\ref{gij_down}) and (\ref{Kij_up}) do not fulfill the
constraint equations of general relativity. In order to find a
3-metric and extrinsic curvature which do fulfill the constraints, we
now apply the York procedure to project the PN 3-metric and extrinsic
curvature onto the solution manifold of general relativity.  In this
procedure we freely specify a 3-metric $\bar{g}_{ij}$, a symmetric
traceless tensor $\bar{A}^{ij}$ and a scalar $K$.  We then solve the
constraint equations
\begin{eqnarray}
\label{york_ham}
0 &=& 
  \bar{\nabla}^2 \Psi -\frac{1}{8} \Psi \bar{R} 
  -\frac{1}{12}  \Psi^5 K^2    \nonumber \\
&&+\frac{1}{8} \Psi^{-7} 
  (\bar{A}^{ij} +\bar{L}W^{ij} )
  (\bar{A}^{kl} + \bar{L}W^{kl})\bar{g}_{ik}\bar{g}_{jl}  
\end{eqnarray}
and
\begin{equation}
\label{york_mom}
0= \bar{\Delta}_L W^i -\frac{2}{3} \Psi^6 \bar{\nabla}^i K 
   +\bar{\nabla}_j \bar{A}^{ij} 
\end{equation}  
for $\Psi$ and $W^i$. Here $\bar{\nabla}$ and $\bar{R}$ are the covariant
derivative and Ricci scalar associated with the 3-metric
$\bar{g}_{ij}$, 
$\bar{L}W^{ij} = \bar{\nabla}^iW^j + \bar{\nabla}^jW^i - 
 \frac{2}{3}\bar{g}^{ij}\bar{\nabla}_kW^k$, and 
$\bar{\Delta}_L W^i = \bar{\nabla}_j\bar{L}W^{ij}$.
Then
\begin{equation}
g_{ij} = \Psi^4 \bar{g}_{ij} 
\end{equation}  
and 
\begin{equation}
K^{ij}= \Psi^{-10}(\bar{A}^{ij} +\bar{L}W^{ij})+\frac{1}{3} g^{ij} K
\end{equation} 
with $g^{ij}$ being the inverse of $g_{ij}$ will
satisfy the constraints of general relativity.

\subsection{Application of the York Procedure to the PN data}

The idea is to base the freely specifiable quantities $\bar{g}_{ij}$,
$\bar{A}^{ij}$, and $K$ on the
PN 3-metric, the traceless part of the PN extrinsic curvature
and the trace of the PN extrinsic curvature.
The specific PN expressions we use are
\begin{equation}
g^{5}_{ij}
=\psi_{5}^{4} \delta_{ij} + 
\ftimes\left(
 h_{ij (4)}^{TT} + h_{ij (5)}^{TT} 
       \right) 
\end{equation}
and 
\begin{eqnarray}
K_{5}^{ij}
= -\psi_{5}^{-10}\left[
        \tilde{\pi}^{ij}_{(3)} + \frac{1}{2}\ftimes\dot{h}_{ij (4)}^{TT}
        + \ftimes(\phi_{(2)}\tilde{\pi}^{ij}_{(3)})^{TT}  \right]
\end{eqnarray}
with 
\begin{eqnarray}
\psi_{5} &=& 
1 
+\frac{1}{2r_1} \left(m_1+\frac{p_{1}^2}{2m_{1}}-\frac{m_1m_2}{2r_{12}}\right)
\nonumber \\ 
&& \,\,\,\,\,\, 
+\frac{1}{2r_2} \left(m_2+\frac{p_{2}^2}{2m_{2}}-\frac{m_1m_2}{2r_{12}}\right).
\end{eqnarray}
Here $g_{5}^{ij}$, $K_{5}^{ij}$ and $\psi_{5}$ are the
PN expressions (\ref{gij_down}), (\ref{Kij_up}) and (\ref{psiPN})
with all terms of $O(\epsilon^6)$ or higher dropped. 
% The additional
% factor $f$ in $g_{5}^{ij}$ and $K_{5}^{ij}$ is an attenuation factor given 
% by either by 
% $ f = 1 $ 
% or
%       % f = 0.5*( 1 + Tanh[a( 1/x - (b^2)*x )] )       
%       % x = (phi2)/4
% \begin{equation}
% \label{AttFac}
% f =  0.5  \left\{ 1 + \tanh[a( 4/ \phi_{(2)} - b^2 \phi_{(2)}/4 )] \right\}, 
% \end{equation}
% with 
% $  \phi_{(2)} = \frac{ 4 m_1 }{r_1} +  \frac{ 4 m_2 }{r_2}$
% and where $a$ and $b$ are constants of order $1$.
% The factor in Eq.~(\ref{AttFac}) is constructed such 
% that it completely disappears if all 
% the $\epsilon$'s are reinserted and everything is expanded up to
% $\epsilon^5$. 
% By adjusting the constants $a$ and $b$ we choose $f$
% such that it is zero at the singularities and one elsewhere. 
% This causes $g^{5}_{ij}$ at the  singularities to be conformally flat
% and $K_{5}^{ij}$ to be of Bowen-York form, i.e.\ at the singularities 
% we be have pure punctures.

For $\bar{g}_{ij}$ we choose the conformally rescaled metric 
\begin{equation}
\bar{g}_{ij} = \psi_{5}^{-4} g^{5}_{ij}
= \delta_{ij} 
 + \ftimes \psi_{5}^{-4} \left(h_{ij (4)}^{TT} + h_{ij (5)}^{TT} \right) ,
\end{equation}
which has the advantage of being regular near the black holes. We also
conformally rescale the extrinsic curvature and pick
\begin{eqnarray}
\bar{A}^{ij} &=& \psi_{5}^{10} 
      \left(K_{5}^{ij} -\frac{1}{3} g_{5}^{ij} K_{5} \right)  \nonumber \\
&=& -\tilde{\pi}^{ij}_{(3)} 
                - \ftimes \left[\frac{1}{2}\dot{h}_{ij (4)}^{TT}
                           + (\phi_{(2)}\tilde{\pi}^{ij}_{(3)})^{TT} \right] 
                - \frac{\psi_{5}^{10}}{3} g_{5}^{ij} K_{5} ,   \nonumber \\
\end{eqnarray}
where $K_{5} =  g^{5}_{ij} K_{5}^{ij}$.
Finally, since we only consider terms up to order $\epsilon^5$ and because
$K_{PN} = O(\epsilon^7)$ we choose
\begin{equation}
K=0 .
\end{equation}

The metric $\bar{g}_{ij}$ is regular
near the black holes. If $r_A$ denotes the distance to the
singularity, we have
\begin{equation}
\label{psi_5_scal}
\psi_{5} \sim O(1/r_A)
\end{equation} 
and 
$h_{ij (4)}^{TT} + h_{ij (5)}^{TT} \sim 1/r_A$ 
so that
\begin{equation}
\bar{g}_{ij}  \sim \delta_{ij} + \ftimes O(r_A^3) .
\end{equation}
This means that Christoffel symbols and Ricci scalar computed from the 
3-metric $\bar{g}_{ij}$ go as
\begin{equation}
\label{Gamma_scal}
\bar{\Gamma}^{k}_{ij}  \sim \ftimes O(r_A^2) .
\end{equation}
and
\begin{equation}
\label{Rbar_scal}
\bar{R}  \sim  \ftimes O(r_A) .
\end{equation}
%Since $\tilde{\pi}^{ij}_{(3)}  \sim \ftimes O(1/r_A^2)$ 
We also have
\begin{equation}
K_{5}^{ij} \sim  O(r_A^8) + \ftimes O(r_A^7)
\end{equation} 
and thus
\begin{equation}
\label{K_scal}  
K_{5} \sim \ftimes O(r_A^4) + \ftimes O(r_A^3)
\end{equation}  
and
\begin{equation}
\label{Aij_scal}  
\bar{A}^{ij} \sim O(1/r_A^2) + \ftimes O(1/r_A^3).
\end{equation}  
So except for $\bar{A}^{ij}$ and $\psi_5$ all quantities are well
behaved near the black holes.

The remaining problem is to solve (\ref{york_ham}) and (\ref{york_mom})
numerically. Since the PN metric is an approximate solution it is clear
that $\Psi \approx \psi_5$ and hence that $\Psi$ will diverge near the black
hole, which of course is problematic 
when $\bar{\nabla}^2 \Psi \sim O(1/r^3)$ is calculated by finite differencing
in numeric computations.
In order to overcome this problem we make the ansatz
\begin{equation}
\label{PsiAnsatz}
\Psi = \psi_{5} + u ,
\end{equation}
which in the case of the original puncture data suffices to regularize
the constraint equations \cite{Brandt97b}.  With this ansatz
Eq.~(\ref{york_ham}) becomes
\begin{eqnarray}
\label{ham}
0 &=& 
  \bar{\nabla}^2 u + (\bar{g}^{ij}-\delta^{ij})\partial_i \partial_j \psi_{5} 
  - \bar{g}^{ij}  \bar{\Gamma}^{k}_{ij} \partial_k \psi_{5}  
  -\frac{1}{8} \Psi \bar{R}   \nonumber  \\
&&
%%-\frac{1}{12}  \Psi^5 K^2   
  +\frac{1}{8} \Psi^{-7} 
  (\bar{A}^{ij} + \bar{L}W^{ij} )
  (\bar{A}^{kl} + \bar{L}W^{kl})\bar{g}_{ik}\bar{g}_{jl} ,
\end{eqnarray}
where the term
\begin{equation}
  \delta^{ij} \partial_i \partial_j \psi_{5} = 0
\end{equation}
has been subtracted. This term vanishes analytically away from the punctures
and it is numerically advantageous to use it to cancel the corresponding term
in $\bar{g}^{ij}$.
Using Eqs.~(\ref{psi_5_scal}), (\ref{Gamma_scal}),
(\ref{Rbar_scal}) and (\ref{Aij_scal}) one can check that all terms in
Eq.~(\ref{ham}) are finite.
Furthermore we split $\bar{A}^{ij}$ into the two parts
\begin{equation}
\bar{A}_S^{ij}= 
 -\tilde{\pi}^{ij}_{(3)} 
 - \frac{1}{2}\dot{h}_{ij (4)}^{TT}
 - (\phi_{(2)}\tilde{\pi}^{ij}_{(3)})^{TT}
\end{equation}
and
\begin{equation}
\bar{A}_R^{ij}= \bar{A}^{ij} - \bar{A}_S^{ij} 
%=
%% - \ftimes \left[ \frac{1}{2}\dot{h}_{ij (4)}^{TT}
%%                 + (\phi_{(2)}\tilde{\pi}^{ij}_{(3)})^{TT} \right]
% - \frac{\psi_{5}^{10}}{3} g_{5}^{ij} K_{5} ,  
\end{equation}
so that $\bar{A}^{ij}= \bar{A}_S^{ij} + \bar{A}_R^{ij}$ .
The advantage of splitting $\bar{A}^{ij}$ in this way is that, analytically,
\begin{equation}
\label{DivA_Sij}
\partial_j \bar{A}_S^{ij}= 0 
\end{equation}
away from the punctures.
Using Eq.~(\ref{DivA_Sij}) the constraint equation (\ref{york_mom}) 
simplifies to
\begin{equation}
\label{mom}
\bar{\Delta}_L W^i 
%%-\frac{2}{3} \Psi^6 \bar{\nabla}^i K
+\bar{\Gamma}^{i}_{jk} \bar{A}_S^{kj} +\bar{\Gamma}^{j}_{jk} \bar{A}_S^{ik}
+\bar{\nabla}_j \bar{A}_R^{ij} =0 . 
\end{equation}
Eqs.~(\ref{ham}) and (\ref{mom}) now can be solved numerically for $u$
and $W^i$ given the boundary conditions that $u\rightarrow 0$ and
$W^i\rightarrow 0$ for $r\rightarrow\infty$. There are no additional
boundary conditions at the punctures, rather we assume that there
exists a unique solution for which $u$ and $W^i$ are $C^2$ at the punctures,
which has been proven to be the case for the simpler example considered
in \cite{Brandt97b}.

\subsection{Ambiguities in the application of the York procedure}
\label{subsec_YorkAmbig}

Note that the York procedure explained above 
was applied to the conformally rescaled quantities
$\bar{g}_{ij}$ and $\bar{A}^{ij}$. There is a priori no reason
for using $\bar{g}_{ij}$ and $\bar{A}^{ij}$. In principle
we could have also started directly with $g^{PN}_{ij}$ and $ K_{PN}^{ij}$ or
with $g^{PN}_{ij}$ and $ K_{PN}^{ij}$ scaled by any function $\Omega$, i.e.
with
\begin{equation}
\tilde{g}^{PN}_{ij} = \Omega^{4}  g^{PN}_{ij}
\end{equation}
\begin{equation}
\tilde{K}_{PN}^{ij} = \Omega^{-10} K_{PN}^{ij}
\end{equation}
and the York procedure would still yield a solution to the constraints.
Each of these different starting points will in general yield different 
results for $g_{ij}$ and $K_{ij}$ depending on $\Omega$. The solution for
$g_{ij}$ and $K_{ij}$ becomes independent of $\Omega$ only if $K_{5}^{ij}$
already fulfills the momentum constraint, which is not the case for the PN
expressions.
As an example of this freedom we expand $\Omega$ in $\epsilon$ and choose
\begin{equation}
\label{Omega-choice}
\Omega = 1 + \epsilon^4 Q + O(\epsilon^6) .
\end{equation}
Due to the absence 
of $O(\epsilon^2)$ terms in $\Omega$ we obtain the simple result
\begin{eqnarray}
\tilde{g}^{PN}_{ij} 
&=& (1 +\epsilon^4 Q +O(\epsilon^6))^{4} g^{PN}_{ij} \nonumber \\
&=&(\psi_{PN} +\epsilon^4 Q )^4   \delta_{ij}   \nonumber \\
&&+ \epsilon^4 h_{(4) ij }^{TT} + \epsilon^5 h_{(5) ij }^{TT} +O(\epsilon^6)
\end{eqnarray}
and
\begin{eqnarray}
\tilde{K}_{PN}^{ij} 
&=& (1+\epsilon^4 Q +O(\epsilon^6) )^{-10} K_{PN}^{ij} \nonumber \\
&=& -(\psi_{PN} +\epsilon^4 Q)^{-10} 
\Big[ \epsilon^3 \tilde{\pi}^{ij}_{(3)} \nonumber \\
&&              + \epsilon^5 \frac{1}{2}\dot{h}_{ij (4)}^{TT}
                + \epsilon^5 (\phi_{(2)}\tilde{\pi}^{ij}_{(3)})^{TT}
                \Big] + O(\epsilon^6) .
\end{eqnarray}
We see that $g^{PN}_{ij}$ and $K_{PN}^{ij}$ differ from 
$\tilde{g}^{PN}_{ij}$ and $\tilde{K}_{PN}^{ij}$ only in the factor
\begin{equation}
\label{psi-shift}
\tilde{\psi}_{PN} = \psi_{PN} +\epsilon^4 Q .  
\end{equation}  
This shows that an overall conformal rescaling by 
$\Omega = 1+\epsilon^4 Q$
can be understood as a shift (by $\epsilon^4 Q$) in the PN conformal factor.

Furthermore note that any 3-metric $g_{ij}$ and extrinsic curvature
$K_{ij}$ constructed by the method explained above are in general
different from the PN expressions for 3-metric and extrinsic
curvature.  If one assumes that the PN expressions are valid and thus
astrophysically realistic (at least in a certain regime), one can aim
to minimize the difference between $g_{ij}$ and $K_{ij}$ and the PN
expressions in this regime.  We will later show that the scaling in
Eq.~(\ref{Omega-choice}) can be used to improve $g_{ij}$ such that the
ADM mass of the system after the York procedure is close to what is
predicted by pure PN theory in the regime where PN theory is valid.

\section{Numerics}
\label{Numerics}

We now demonstrate that our method for solving the constraints in
Eqs.~(\ref{ham}) and (\ref{mom}) leads to convergent numerical
solutions. We use second order finite differencing together with a
multigrid elliptic solver (BAM\_Elliptic in
Cactus~\cite{Bruegmann99b}). All grids have uniform resolution.  The
two black hole punctures are always staggered between grid points on
the finest grid in the multigrid scheme.  Since we absorb all
diverging terms in the conformal factor the solutions $u$ and $W^i$ of
Eqs.~(\ref{ham}) and (\ref{mom}) are regular everywhere, so that no
black hole excision or inner boundary conditions are needed.  As outer
boundary conditions we use Robin conditions, i.e.\ we assume that $u
\propto 1/r$ and $W^i \propto 1/r$, where $r$ is the distance to the
center of mass. In the case of the vector potential this is a
simplifying assumption that works reasonably well in practice.

%%%%%%%%%%%%%%%%%%%%%%%%%%%%%%%%%%%%%%%%%%%%%%%%%%%%%%%%%%%%%%%%%%%%%%%%%%
% begin FIGURES
%%%%%%%%%%%%%%%%%%%%%%%%%%%%%%%%%%%%%%%%%%%%%%%%%%%%%%%%%%%%%%%%%%%%%%%%%%
\begin{figure}
%\centerline{\psfig{gnupl.ps,height=5cm,width=5cm}}
\epsfxsize=8.5cm 
\epsfbox{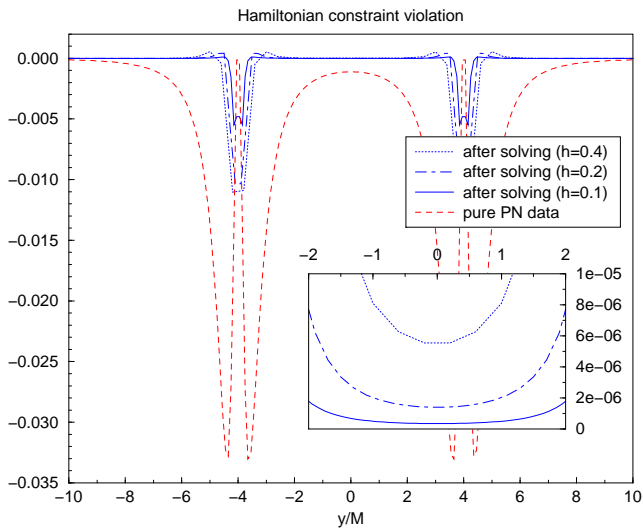}
\vspace{.4cm}
\caption{Hamiltonian constraint violation for a black hole 
separation of $r_{12}=8M$. The Hamiltonian constraint of 
pure PN data is much larger than the
Hamiltonian constraint after solving (i.e.\ applying the 
the York procedure). We numerically solve for three different resolutions
$h$. The inset is a blow up of the central region, which shows that our
numerical scheme is second order convergent as expected.
}
\label{fig_ham}
\end{figure}

\begin{figure}
%\centerline{\psfig{gnupl.ps,height=5cm,width=5cm}}
\epsfxsize=8.5cm 
\epsfbox{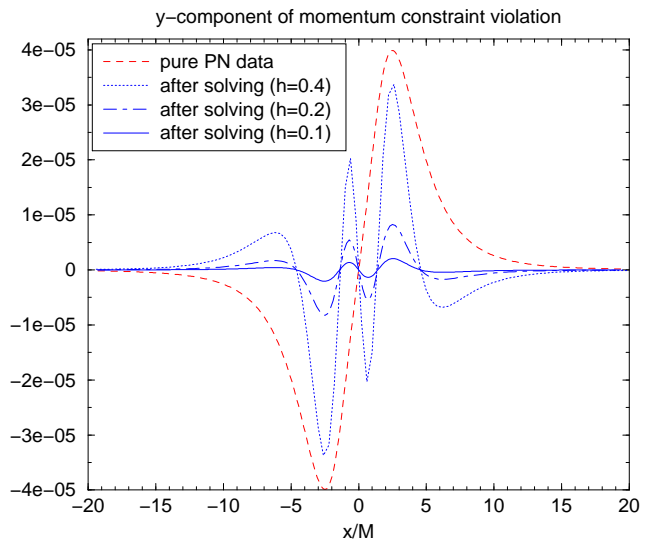}
\vspace{.4cm}
\caption{    
The momentum constraint for a separation of $r_{12}=8M$. We observe second
order convergence in the resolution $h$ after solving. 
The momentum constraint violation of pure
PN data is larger than after solving.
}
\label{momy}
\end{figure}

\begin{figure}
%\centerline{\psfig{gnupl.ps,height=5cm,width=5cm}}
\epsfxsize=8.5cm 
\epsfbox{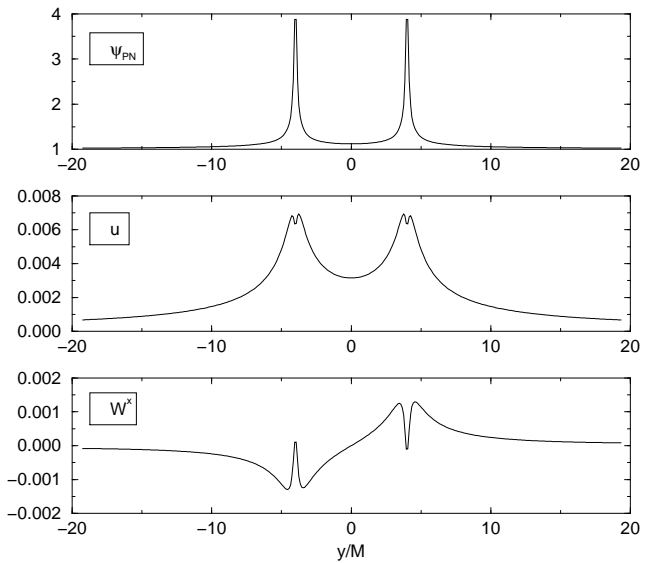}
\vspace{.4cm}
\caption{  
The solutions of $u$ and $W^x$ along the y-axis for a black hole separation 
of $r_{12}=8M$.  For comparison we also show $\psi_{PN}$, which diverges
at $y=\pm 4$. 
}
\label{psi_phi_wx} 
\end{figure}

\begin{figure}
%\centerline{\psfig{gnupl.ps,height=5cm,width=5cm}}
\epsfxsize=9cm
\epsfbox{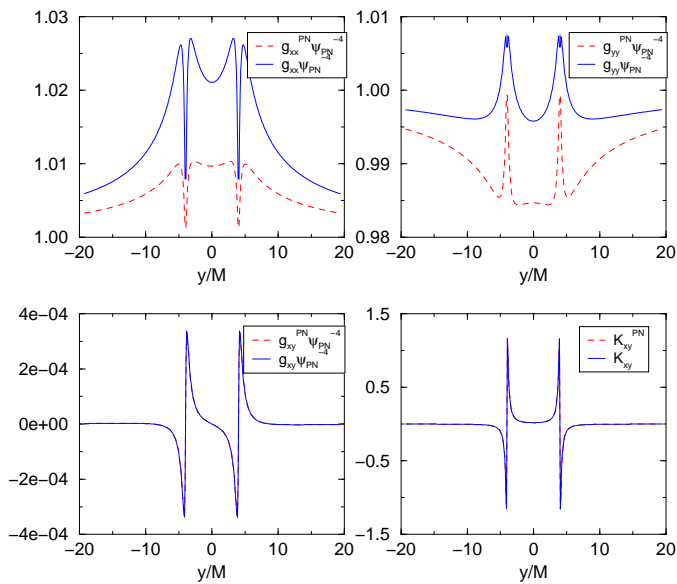}
\vspace{.4cm}
\caption{
Components of the 3-metric and extrinsic curvature for a black hole separation
of $r_{12}=8M$. The data are shown before (dashed lines) and after applying
the York procedure (solid lines).
The components of the 3-metric change on the order of $\sim 1\%$.
}           
\label{gxx_gyy_gxy_kxy} 
\end{figure}

\begin{figure}
%\centerline{\psfig{gnupl.ps,height=5cm,width=5cm}}
\epsfxsize=8.5cm
\epsfbox{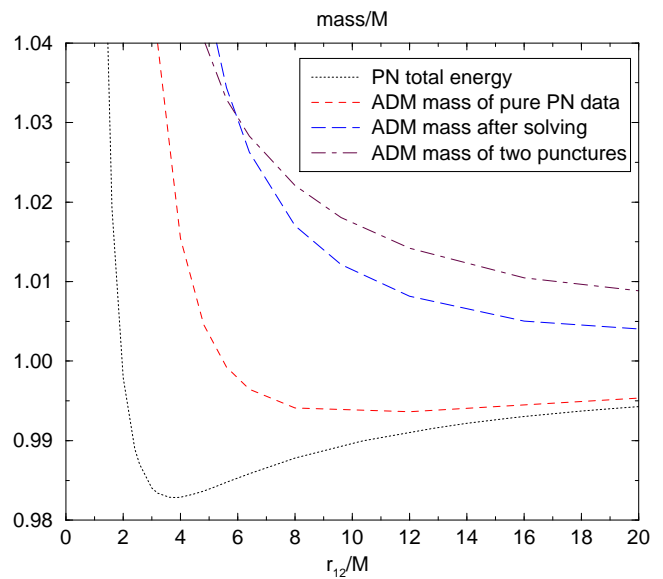}
\vspace{.4cm}
\caption{
PN energy of Eq.~(\ref{PN_Etot}) and ADM masses before and after
solving (i.e.\ applying the York procedure) 
versus coordinate separation $r_{12}$ along the PN inspiral sequence.
The data here were computed by keeping all momentum terms
in $\psi_{PN}$, without consistently dropping higher order terms.
In this case the ADM mass of pure PN data does not agree well with 
the PN energy. The ADM mass after solving (with $q=0.0$) 
increases on the order of $\sim 1\%$, when compared 
to the ADM mass of pure PN data. Furthermore the ADM mass after solving
increases with decreasing separation, which is physically not acceptable.
For comparison we also show the ADM mass of two puncture black holes along
the PN sequence with constant bare masses, which show a similar increase
in ADM mass.
}           
\label{energies1} 
\end{figure}

\begin{figure}
%\centerline{\psfig{gnupl.ps,height=5cm,width=5cm}}
\epsfxsize=8.5cm
\epsfbox{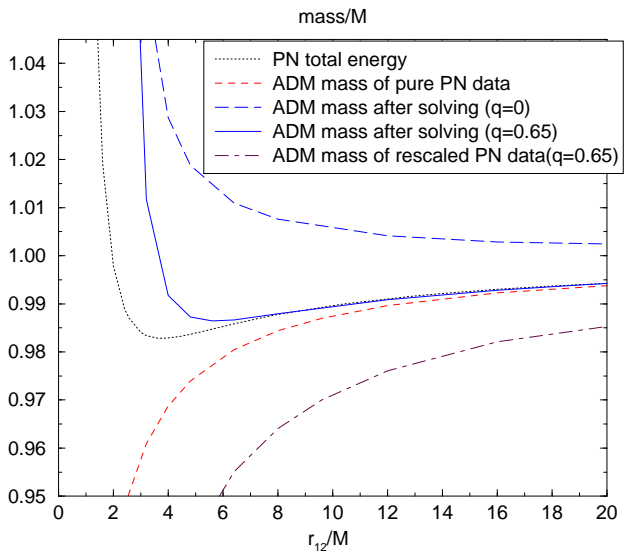}
\vspace{.4cm}
\caption{           
PN energy of Eq.~(\ref{PN_Etot}) and ADM masses versus coordinate separation
$r_{12}$ along the PN inspiral sequence.
Shown are the ADM masses
before and after applying the York procedure with both 
$q=0$ and $q=0.65$.
Here all data are computed by consistently keeping momentum terms 
in $\psi_{PN}$ only up to Newtonian order.
The ADM mass of pure PN data now agrees better with 
the PN energy. The York procedure with $q=0.0$ 
again increases the ADM mass on the order of $\sim 1\%$, when compared 
to the ADM mass of pure PN data. The ADM mass after solving with 
$q=0.65$, however, does not change very much and it also closely follows
the PN energy down to $r_{12} \approx 6M$. Furthermore until
$r_{12} \approx 5.6M$ it is physically reasonable since it decreases with
decreasing separation. For comparison we also show the ADM mass curve
of rescaled PN data (with $q=0.65$). These data, however, have no direct
physical significance.
}           
\label{energies2} 
\end{figure}

\begin{figure}
%\centerline{\psfig{gnupl.ps,height=5cm,width=5cm}}
\epsfxsize=8.5cm
\epsfbox{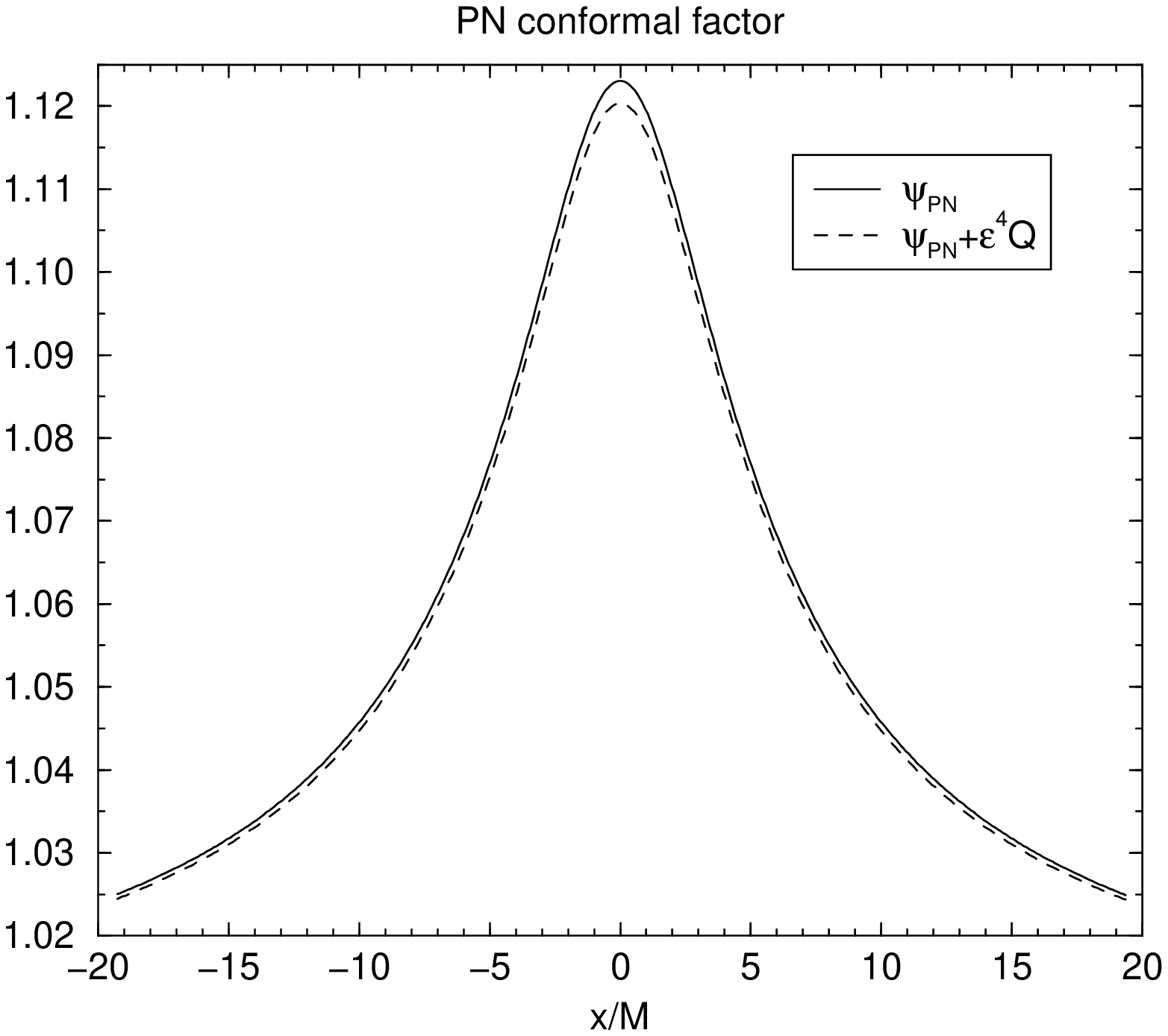}
\vspace{.4cm}
\caption{The conformal factors $\psi_{PN}$ and 
$\tilde{\psi}_{PN} = \psi_{PN}+\epsilon^4 Q$,
before and after rescaling with $q=0.65$ for $r_{12}=8M$.
The difference between $\psi_{PN}$ and $\tilde{\psi}_{PN}$ is small.
}           
\label{psi_EpCorr1} 
\end{figure}

%%%%%%%%%%%%%%%%%%%%%%%%%%%%%%%%%%%%%%%%%%%%%%%%%%%%%%%%%%%%%%%%%%%%%%%%%%
% end FIGURES
%%%%%%%%%%%%%%%%%%%%%%%%%%%%%%%%%%%%%%%%%%%%%%%%%%%%%%%%%%%%%%%%%%%%%%%%%%

For the numerical work in this paper we consider non-spinning
equal mass binaries with their center of mass at rest at the
origin. The binaries are in quasi-circular orbits in the sense that we
use Eq.~(\ref{p_circ}) to set the momentum of the two black holes
before solving the constraints.  The two black holes are on the
y-axis, such that their momenta point in the positive and negative
x-directions, resulting in an angular momentum along the z-direction.
Fig.~\ref{fig_ham} shows the Hamiltonian constraint violation of pure
PN data (dashed line), i.e.\ before solving the constraints, as well
as the Hamiltonian constraint after solving at three different
resolutions $h$. After the elliptic solve the constraint equations
(\ref{ham}) and (\ref{mom}) are satisfied to within a given tolerance
of $10^{-10}$ in the l2-norm, but to study convergence we show the ADM
constraints computed from $g_{ij}$ and $K_{ij}$. The two black holes
are at $y=\pm 4$.  One can see that the constraint violation after the
York procedure is much smaller than the constraint violation of pure
PN data.  The inset in Fig.~\ref{fig_ham} is a blow up of the center
and shows second order convergence to zero in the Hamiltonian
constraint after solving.  We also observe second order convergence to
zero in the momentum constraint. As an example we show the y-component
of the momentum constraint in Fig.~\ref{momy}. We see that pure PN
data violates the constraints. In Fig.~\ref{psi_phi_wx} we plot the
solutions $u$ and $W^x$ along the y-axis, which contains the black
holes. As expected they are regular, unlike $\psi_{PN}$ which diverges
at the black hole locations of $y=\pm 4$.

As expected, after applying the York procedure $g_{ij}$ and $K^{ij}$
are different from the pure PN expressions $g^{PN}_{ij}$ and 
$K_{PN}^{ij}$.  Fig.~\ref{gxx_gyy_gxy_kxy} shows a comparison of
several components of the 3-metrics $\psi_{PN}^{-4} g_{ij}$ and
$\psi_{PN}^{-4} g^{PN}_{ij}$.  As one can see, the components of
$g_{ij}$ exhibit an increase on the order of 
$\sim1\%$ when compared to $g^{PN}_{ij}$.  
\begin{table}
\caption{Selected components of the 3-metric, extrinsic curvature 
and $h_{ij (4)}^{TT}$ at the
point $x=0$, $y=12.2M$, $z=0$ for two black holes located on the 
y-axis at $y=\pm 5.2M$. 
The change in the 3-metric induced by solving the constraints
without first rescaling $\psi_{PN}$  
has about the same magnitude as the PN corrections at $O(\epsilon^4)$.
The data here are computed by inconsistently keeping all 
higher order momentum terms in $\psi_{PN}$.
\label{PN_vs_noEpcorr-table}}
\begin{tabular}{|c|c|c|}
\hline
 PN value                & Value after     &  relative   \\
 (up to $O(\epsilon^5)$) & solving ($q=0$) &  difference \\
\hline
$g^{PN}_{xx}=1.21866$ & $g_{xx}=1.22285$   
                        & $\frac{g_{xx}-g^{PN}_{xx}}{g^{PN}_{xx}}=0.0034$ \\
%$K^{PN}_{xy}=-2.2341\times 10^{-3}$ & $K_{xy}=-2.2617\times 10^{-3}$ 
% & $\frac{K_{xy}-K^{PN}_{xy}}{K^{PN}_{xy}}=-0.0124$ \\
$K^{PN}_{xy}=-0.0022341$ & $K_{xy}=-0.0022617$
                & $\frac{K_{xy}-K^{PN}_{xy}}{K^{PN}_{xy}}=-0.012$ \\
\hline
 PN metric               & TT term in metric  &  relative size of      \\
 (up to $O(\epsilon^5)$) & of $O(\epsilon^4)$ &  $O(\epsilon^4)$ correction\\
\hline
$g^{PN}_{xx}=1.21866$ & $h_{xx (4)}^{TT}=0.00443$
 & $\frac{h_{xx (4)}^{TT}}{g^{PN}_{xx}}=0.0036$ \\
\hline
\end{tabular}
\end{table}
The same conclusion is reached by looking at
Tab.~\ref{PN_vs_noEpcorr-table}, which shows the 3-metric and
extrinsic curvature before and after applying the York procedure.
Furthermore Tab.~\ref{PN_vs_noEpcorr-table} shows that the increase in
the 3-metric due to applying the York procedure has about the
same order of magnitude as the PN corrections at $O(\epsilon^4)$.
Since this happens in a region far enough from the particles that PN
theory can actually be trusted to give realistic values, it means that
solving the elliptic equations introduces significant differences
between $g_{ij}$ and $g^{PN}_{ij}$ in the outer region due to changes
in the inner region.  Before we suggest how this problem can be
addressed, let us also consider the ADM mass of the system, which is a
coordinate invariant quantity.

We compute the ADM mass along PN inspiral sequences constructed from
PN circular orbits with different radii. Along such a sequence the
bare masses $m_1$ and $m_2$ are kept constant and the momenta are
computed from Eq.~(\ref{p_circ}) for circular
orbits. Fig.~\ref{energies1} shows the numerically computed ADM mass
of pure PN initial data (dashed line), the ADM mass of the data
obtained after applying the York procedure (long dashed line), as well
as the PN total energy (dotted line) of Eq.~(\ref{PN_Etot}). In 
Fig.~\ref{energies1} and the following figures we plot data for
$r_{12}$ between $1$ and $20M$. But note that it has to be expected
that the PN data becomes inaccurate for small $r_{12}$, for example 
for $r_{12} \approx 4M$ where the black holes are close to the
fiducial ISCO of the PN data. 

In Fig.~\ref{energies1}, we again observe an increase of $\sim 1\%$ in
the ADM mass after applying the York procedure.  A further problem is
that none of the numerically determined ADM masses in
Fig.~\ref{energies1} agrees very well with the PN energy
(\ref{PN_Etot}).  This problem stems from the fact that the PN initial
data in Fig.~\ref{energies1} have been obtained by inserting the
momentum (\ref{p_circ}) as it is into the expressions for 3-metric and
extrinsic curvature of Sec.~\ref{sec-expr_metric_curv} without
consistently dropping terms of $O(\epsilon^6)$ or higher. Since all PN
corrections to the momentum are positive, the main effect of this
inconsistency is to increase $\psi_{PN}$ given by Eqs.~(\ref{psiPN})
and (\ref{EnergyInPsi}).  The result is that the numerically computed
ADM masses before and after applying the York procedure show
physically unacceptable behavior: (i) the ADM mass of pure PN data
approaches the PN energy (\ref{PN_Etot}) only very slowly at large
separations, and (ii) the ADM mass of the data after applying the York
procedure monotonically increases with decreasing separation. This is
physically not reasonable because the system is supposed to loose
energy due to the emission of gravitational radiation.  For reference
the ADM mass (dot dashed line) for a sequence of two black hole
punctures with constant bare masses and with the same PN momentum
(\ref{p_circ}) is also shown in Fig.~\ref{energies1}. Along this
sequence the ADM mass of the punctures also unphysically rises with
decreasing separation, which is not surprising since the assumption of
constant bare masses for punctures ignores the growing contribution of
$u$ to the conformal factor with decreasing separation of the
punctures.  In all cases studied by us the solution $u$ of
Eq.~(\ref{ham}) is indeed positive, which translates directly into an
increase in the mass.

%, so that the overall conformal factor $\Psi = \psi_{PN} + u $
%is larger than $\psi_{PN}$. This increase in the conformal factor
%directly translates into an increase in mass.  

Of course, the question is how we can improve our data so that its
behavior is physically more realistic. One can argue that part of
the additional energy is tied to an increased local mass of the
individual black holes. In fact, for constant bare masses there is a
strong growth in the apparent horizon masses. A standard approach is
therefore to rescale the bare masses to keep the apparent horizon mass
fixed and to define a binding energy by subtracting the
apparent horizon masses from the total mass, e.g.\ \cite{Cook94}.
We plan to compute apparent horizon masses for a future publication. 
However, in general it is not possible to unambiguously
define a local mass for general relativistic data, and the accuracy
and validity of the estimate for the binding energy therefore depends
on, for example, how close the black holes are.

As an alternative we have experimented here with a mass correction
that is tied to properties of the PN approximation. As a first step
let us keep momentum terms of Eq.~(\ref{p_circ}) in the PN conformal
factor $\psi_{PN}$ (see Eqs.~(\ref{psiPN}) and (\ref{EnergyInPsi}))
only up to the appropriate order and to consistently drop all terms of
$O(\epsilon^6)$ and higher. This amounts to just using the first
Newtonian term of the momentum (\ref{p_circ}) in $\psi_{PN}$.  The
results are shown in Fig.~\ref{energies2}. The ADM mass of pure PN
data (dashed line) now much better approaches the PN energy for large
separations.  Yet, the ADM mass after simply applying the York
procedure (long dashed line) still shows an increase of order $\sim
1\%$ when compared to pure PN data.  
If we want more
physical mass curves we have to prevent this increase by preventing the
increase in the conformal
factor. We will take advantage of the freedom in the York
procedure mentioned in Sec.~\ref{subsec_YorkAmbig} and use
the conformal rescaling of Eq.~(\ref{Omega-choice}) before applying
the York procedure.  From Eq.~(\ref{psi-shift}) we see that then the
overall conformal factor becomes
\begin{equation}
\Psi = \tilde{\psi}_{PN} + u  = \psi_{PN} +\epsilon^4 Q + u .
\end{equation}
Hence, if we choose an appropriate $Q$, we have a chance of compensating
$u$ such that $\Psi \approx \psi_{PN}$ at least in the region far from 
the black holes where PN theory is valid.

Now, in the limit of $r_{12} \rightarrow \infty$
the pure PN data we use as a starting point represent
two Schwarzschild black holes at rest (in isotropic coordinates).
Thus $u$ is zero for infinite separation and 
we therefore expect that $u$ goes like
$u \propto 1/r_{12}$ for large $r_{12}$.                        
On the other hand we also have  $u \propto 1/r$ 
so that we expect that $u$ is well approximated by
\begin{equation}
u  \approx \frac{N}{r_{12}r} 
\end{equation}
for large $r$, where $N$ is some numerical constant.
Since we want $Q +u \approx 0$, we choose
\begin{equation}
\label{Q_choice}
Q= -q\frac{m_1 m_2}{2r_{12}}\left(\frac{1}{2r_{1}} +\frac{1}{2r_{2}}\right).
\end{equation}
Here $q$ is a free parameter, which has to be chosen such that 
$Q+u \approx 0$ for large separations.

It turns out that for $q=0.65$ we get physically more reasonable mass
curves in the regime where PN theory is expected to be valid.
The solid line in Fig.~\ref{energies2} shows the ADM mass obtained for
different separations if we apply the following extended York procedure:
(i)   start with the pure PN initial data,
(ii)  rescale $\psi_{PN}$ using Eqs.~(\ref{psi-shift}) and (\ref{Q_choice})
      with $q=0.65$ and
(iii) apply the standard York procedure to the rescaled quantities.
As we can see the ADM mass (solid line) closely follows 
the PN energy (dotted line) in the region 
where we expect PN theory to be valid.
Furthermore for separations greater than $r_{12}\approx 5.6 M$ the ADM
mass decreases with decreasing separation as it should. For smaller
separations the ADM mass again increases. 
In the literature this minimum has often been interpreted
as the location of the innermost stable
circular orbit (ISCO). Note, however, that the PN expressions which we used
up to $O(\epsilon^5)$ are probably close to breaking down around 
$M/r_{12}=1/5.6 \approx 0.2$, so that the ISCO location 
may not be very accurate.
Also the location of the minimum can be shifted if we use
higher order terms in the rescaling of $\psi_{PN}$, i.e.\ if we use 
\begin{equation}
\tilde{\psi}_{PN}  = \psi_{PN}+\epsilon^4 Q + \epsilon^6 Q' .
\end{equation}
The extra $Q'$ term will have no influence in the limit of large distances, 
but it will influence the mass curves at small separation and thus
we can move the minimum. Again one could introduce a one-parameter
family of $Q'$ terms and fit the parameter such that the ADM mass curve
has the minimum at the same place where the PN energy (\ref{PN_Etot})
has a minimum. We decided not to do this since the PN energy
itself may not be very reliable near its minimum.
For comparison, Fig.~\ref{energies2} also shows the ADM mass 
curve (dot dashed line) 
for the PN data rescaled by $Q$ with $q=0.65$, but without applying the
York procedure. This curve has no direct physical meaning, but we
can see that it can be obtained from the curve for pure PN 
data (dashed line) by a downwards shift.
Fig.~\ref{psi_EpCorr1} shows the PN conformal factor before and after
rescaling with $q=0.65$. We see that the change in $\psi_{PN}$ is rather
small.

\begin{figure}
%\centerline{\psfig{gnupl.ps,height=5cm,width=5cm}}
\epsfxsize=8.5cm
\epsfbox{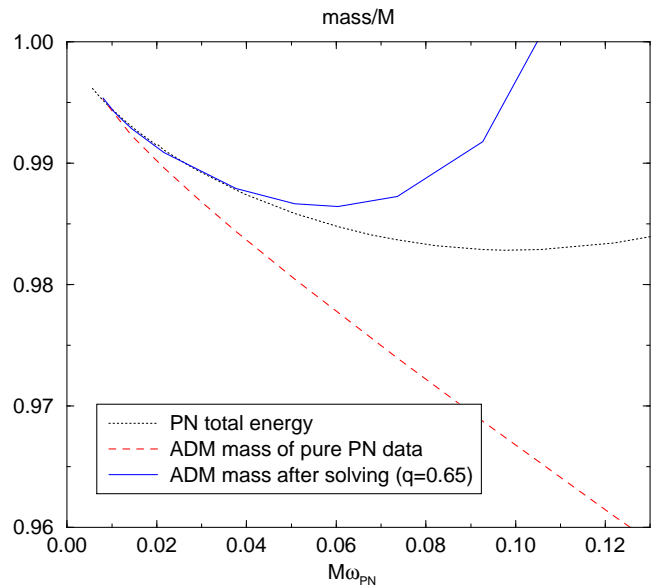}
\vspace{.4cm}
\caption{
PN energy of Eq.~(\ref{PN_Etot}), ADM mass of pure PN data and
ADM mass after solving (with $q=0.65$)
versus the PN angular velocity (\ref{ang_vel}).
The PN energy has a minimum near $M\omega_{PN}\approx 0.1$, which is often
interpreted as the ISCO.
We see that the ADM mass after solving (with $q=0.65$) closely 
follows the PN energy until $M\omega_{PN}\approx 0.05$. Then near
$M\omega_{PN}\approx 0.06$ it has a minimum which 
could be regarded as the ISCO. One has to keep in mind, however, that
the ambiguities in the York procedure in principle allow us to shift 
the location of this minimum. 
}
\label{energies_vs_omega}
\end{figure}

All the masses so far are plotted versus the coordinate separation $r_{12}$.
Fig.~\ref{energies_vs_omega} shows the PN energy (dotted line),
the ADM mass of pure PN data (dashed line) and the ADM mass of data
obtained after rescaling with $q=0.65$ and applying the
York procedure (solid line), versus the PN angular velocity $\omega_{PN}$,
computed for circular orbits from
\begin{eqnarray}
\label{ang_vel}
\left( M \omega_{PN} \right)^2&=&  \frac{64(r_{12}/M)^3}{(1+2r_{12}/M)^6}
+ \frac{\mu}{M} \left( \frac{M}{r_{12}} \right)^4  \nonumber \\
&&+ \left(-\frac{5}{8}\frac{\mu}{M}+\frac{\mu^2}{M^2} \right)
    \left( \frac{M}{r_{12}} \right)^5 .
\end{eqnarray}
Note that $\omega_{PN}$ in Eq.~(\ref{ang_vel}) is written such that
$\omega_{PN}$ is exact up to all PN orders in the limit of $\mu/M
\rightarrow 0$. For $\mu/M > 0$ Eq.~(\ref{ang_vel}) is accurate up to
2PN order. It should be kept in mind, however, that $\omega_{PN}$
probably is not exactly equal to the true angular velocity after
applying the York procedure.  Yet our numerical approach does not
immediately yield an angular velocity which could be used in place of
$\omega_{PN}$.

From Fig.~\ref{energies_vs_omega} we see that the approximate ISCO of
PN theory computed from the 2PN energy is near $M \omega_{PN} = 0.1$,
while the ISCO minimum of our data (after applying the extended York
procedure with $q=0.65$) is near $M \omega_{PN} = 0.06$, which is very
close to the ISCO of test particles in Schwarzschild. Also note that
the ADM mass of pure PN data (dashed line) does not have a minimum at
all.

\begin{table}
\caption{Selected components of the 3-metric, extrinsic curvature 
and $h_{ij (4)}^{TT}$ at the
point $x=0$, $y=12.2M$, $z=0$ for two black holes located on the 
y-axis at $y=\pm 5.2M$. 
The change in the 3-metric induced by solving the constraints
after first rescaling $\psi_{PN}$ (with $q=0.65$) is much smaller 
than the the PN corrections at $O(\epsilon^4)$. The change
in the extrinsic curvature due to solving, however, does not depend much on
$q$ and is about the same whether or not we use the rescaling with $q=0.65$.
Here we have included only Newtonian momentum terms in
$\psi_{PN}$, in order to have a consistent expansion in $\epsilon$.
\label{PN_vs_Epcorr-table}}
\begin{tabular}{|c|c|c|}
\hline
 PN value                & Value after         &  relative   \\
 (up to $O(\epsilon^5)$) & solving ($q=0.65$) &  difference \\
\hline
$g^{PN}_{xx}=1.21738$ & $g_{xx}=1.21783$   
     & $\frac{g_{xx}-g^{PN}_{xx}}{g^{PN}_{xx}}=0.00037$ \\
$K^{PN}_{xy}=-0.0022353$ & $K_{xy}=-0.0022673$   
     & $\frac{K_{xy}-K^{PN}_{xy}}{K^{PN}_{xy}}=-0.014$ \\
\hline
 PN metric               & TT term in metric  &  relative size of      \\
 (up to $O(\epsilon^5)$) & of $O(\epsilon^4)$ &  $O(\epsilon^4)$ correction\\
\hline
$g^{PN}_{xx}=1.21738$ & $h_{xx (4)}^{TT}=0.00443$
 & $\frac{h_{xx (4)}^{TT}}{g^{PN}_{xx}}=0.00364$ \\
\hline
\end{tabular}
\end{table}
In Tab.~\ref{PN_vs_Epcorr-table} we compare some components of
the 3-metric and extrinsic curvature of pure PN data with 
the corresponding quantities obtained after
rescaling with $q=0.65$ and applying the York procedure.
The change in the 3-metric induced by solving the constraints
after first correcting $\psi_{PN}$ (with $q=0.65$) now is much smaller
than the the PN corrections at $O(\epsilon^4)$.
The change in the extrinsic curvature due to solving, however, 
is nearly the same whether or not we use the rescaling with $q=0.65$.

The question arises if the solutions $g_{ij}$ and $K^{ij}$ with
$q=0.65$ are astrophysically more realistic then the pure PN solutions
$g^{PN}_{ij}$ and $K_{PN}^{ij}$. We argue that this is indeed the case
since $g_{ij}$ and $K^{ij}$ with $q=0.65$ are close to $g^{PN}_{ij}$
and $K_{PN}^{ij}$ in the far region where PN is accurate, but in
addition do fulfill the constraint equations of general
relativity. Furthermore the ADM mass curve for $g_{ij}$ and $K^{ij}$
with $q=0.65$ is closer to the PN energy (\ref{PN_Etot}) than the ADM
mass curve of the pure PN solutions $g^{PN}_{ij}$ and $K_{PN}^{ij}$.

\section{Discussion}
\label{Discussion}

For the first time, we have derived fully relativistic black hole
initial data for numerical relativity, starting from 2PN expressions
of the 3-metric and extrinsic curvature in the ADMTT gauge. We have
used the York procedure, and any procedure for projecting the PN data
onto the solution manifold of general relativity will introduce
changes to the PN data. The larger the violation of the constraints by
the PN data, the larger the change in the solution process will be. In
principle one may loose the PN characteristics that distinguished the
PN data from other approaches in the first place.

As we have seen in Sec.~\ref{Numerics}, the size of these changes
depends on how exactly we employ the York procedure for the
projection. We find that the extended York procedure (with $q=0.65$)
yields acceptably small changes, so that if the PN data we started
with are astrophysically realistic, the data after solving the
constraints should still be astrophysically relevant. In particular,
our new PN initial data have the nice property that the 3-metric and
extrinsic curvature approach the corresponding 2PN expressions in the
region where PN theory is valid, providing a natural link to the
early inspiral phase of the binary system.  Furthermore, our approach
leads to an easy numerical implementation with a generalized puncture
method.

We consider this work as a first step towards the construction of
astrophysical initial data based on the PN approximation. Although we
are able to remove some of the inherent ambiguity of the method,
several directions should be explored. Since the PN formalism is
unable to unambiguously provide the full information in the black hole
region, one should examine different ways to introduce
black holes. Furthermore it would seem natural to follow the conformal thin
sandwich approach in order to obtain data that corresponds more
closely to a quasi-equilibrium configuration, although in principle we
rather want data for the appropriate PN inspiral rate than for exactly
circular orbits. Note that after the solution process it is not
known how well the orbital parameters correspond to quasi-circular orbits.
One could use, for example, the effective potential 
method \cite{Cook94} with the new PN based data to determine quasi-circular 
orbits of the two black holes. 

Another direction of research is to improve the PN input to our
method. Even though we can solve the constraints for rather small
separations of the black holes, we cannot trust
the numerical data for arbitrarily small coordinate separation, 
because this is where the PN data we start with is probably unreliable.
We have started with a traditional PN approach
\cite{Jaranowski98a}, but there has been significant progress in
extending the validity of the PN approximation to smaller separations
through resummation techniques \cite{Buonanno00a,Damour_T:98,Damour00a}. 
It is an important issue to study how large an intermediate binary black
hole regime might be, where the PN approximation has broken down but
the separation is still significantly larger than the separation for
an approximate ISCO \cite{Brady_P:98}.

In addition, we want to work with higher order PN approximations. The
explicit regularization for 3PN of \cite{Damour:2001bu} could be used
as a starting point. However, our procedure may have
to be modified because of changes in the conformal factor $\psi_{PN}$.  
Finally, Jaranowski and Sch\"afer~\cite{Jaranowski_Schaefer_priv} have
recently provided us with an expression which includes spin terms at
order $(v/c)^3$ in the PN extrinsic curvature. In future work we
intend to use these terms to add spin to the black holes.

Recall that we have concentrated on the near zone. We plan to replace
the near zone expansion of $h_{ij (4)}^{TT}$ with a globally valid
expression. This could be achieved by solving the 
wave equation determining $h_{ij (4)}^{TT}$ (see e.g.~\cite{Schaefer86}) 
numerically, without any near zone
approximations, which would be natural in a method that resorts to
numerics anyway.  If the PN inspiral trajectory is used in this
calculation, the initial slice of our spacetime will already contain
realistic gravitational waves, with the correct PN phasing. When this
spacetime is then evolved numerically we might eventually be able to
compute numerical wave forms which continuously match PN wave forms.

This brings us to the final goal of our initial data construction,
namely to use it as the starting point for numerical evolutions. As we
pointed out in the introduction, there are now numerical evolution
methods with which we can begin to explore the physical content of any
initial data set by evolution and by extraction of physical quantities
such as detailed wave forms or total radiated energies
\cite{Alcubierre00b,Alcubierre02a,Baker:2002qf}. 
As mentioned in~\cite{Baker:2002qf}, the Lazarus approach provides an
effective method for cross-checking the validity of the results by
choosing different transition times along the binary orbit in the
region where a far limit approximation (such as the PN method) and
full numerical relativity overlap. Only by extending the ability of
full numerical codes to accurately compute several orbits, will we be
able to arrive at a definitive conclusion about the merit of different
initial data sets.

\begin{acknowledgments}
We would like to thank P. Jaranowski and G. Sch\"afer, for many
discussions and sending us their PN expressions in a Mathematica
file. We are also grateful to Dennis Pollney and the Cactus Team for
help on numerical issues related to this work, and to Guillaume Faye
and Carlos O. Lousto who participated in the initial discussions of
this work.  M.C.\ was partially supported by a Marie-Curie Fellowship
(HPMF-CT-1999-00334).  P.D.\ was supported by the EU Programme
`Improving the Human Research Potential and the Socio-Economic
Knowledge Base' (Research Training Network Contract
HPRN-CT-2000-00137). The computations were performed on the SGI Origin
2000 at the Max-Planck-Institut f\"ur Gravitations\-physik and on the
Platinum linux cluster at NCSA.
\end{acknowledgments}

\bibliography{bibtex/references}

\end{document}